\documentclass[aps,pra,twocolumn,superscriptaddress]{revtex4-1}

\usepackage{graphicx}
\usepackage{hyperref}
\usepackage{amsmath}
\usepackage{amssymb}
\usepackage{epstopdf}
\usepackage{multirow}
\usepackage{threeparttable}
\usepackage{lipsum}  
\usepackage[usenames]{color}

\begin{document}

\title{Widely tunable mid-infrared fiber-feedback optical parametric oscillator}
\author{Tingting Yu}
\affiliation{State Key Laboratory of Precision Spectroscopy, East China Normal University, Shanghai 200062, China}

\author{Jianan Fang}
\affiliation{State Key Laboratory of Precision Spectroscopy, East China Normal University, Shanghai 200062, China}

\author{Kun Huang}
\email{khuang@lps.ecnu.edu.cn}
\affiliation{State Key Laboratory of Precision Spectroscopy, East China Normal University, Shanghai 200062, China}
\affiliation{Chongqing Key Laboratory of Precision Optics, Chongqing Institute of East China Normal University, Chongqing 401121, China}
\affiliation{Collaborative Innovation Center of Extreme Optics, Shanxi University, Taiyuan, Shanxi 030006, China}

\author{Heping Zeng}
\email{hpzeng@phy.ecnu.edu.cn}
\affiliation{State Key Laboratory of Precision Spectroscopy, East China Normal University, Shanghai 200062, China}
\affiliation{Chongqing Key Laboratory of Precision Optics, Chongqing Institute of East China Normal University, Chongqing 401121, China}
\affiliation{Chongqing Institute for Brain and Intelligence, Guangyang Bay Laboratory, Chongqing, 400064, China}

\begin{abstract}
Synchronously pumped optical parametric oscillators (OPOs) provide uniquely versatile platforms to generate ultrafast mid-infrared pulses within a spectral range beyond the access of conventional mode-locked lasers. However, conventional OPO sources based on bulk crystals have been plagued by complex optical alignment and large physical footprint. Here, we devise and implement two OPO variants based on a polarization-maintaining fiber-feedback cavity, which allow to robustly deliver sub-picosecond MIR pulses without the need of active stabilization. The first one integrates an erbium-doped fiber into the OPO cavity as the additional gain medium, which significantly reduces the pump threshold and allows stable optical pulse formation within a spectral range of 1553-1586 nm. The second one adopts a chirped poling nonlinear crystal in a passive-fiber cavity to further extend the operation spectral coverage, which facilitates broad tuning ranges of 1350-1768 nm and 2450-4450 nm for the signal and idler bands, respectively. Therefore, the presented mid-infrared OPO source is featured with high compactness, robust operation, and wide tunability, which would be attractive for subsequent applications such as infrared photonics, biomedical examination, and molecular spectroscopy.
\end{abstract}

\maketitle

\section{Introduction}
Mid-infrared (MIR) coherent sources are highly demanded in various applications, such as spectroscopic analysis, material processing, free-space communication, and bio-medical diagnosis \cite{Vodopyanov2020Book, Schliesser2012NP, Hashimoto2019NC}. Nowadays, the MIR light can readily be accessed by quantum cascade lasers (QCLs) that offer a widely tunable range and a high average power \cite{Yao2012NP}. However, the QCLs typically operate in continuous-wave mode, thus requiring sophisticated heat management at room-temperature. Recently, fluoride fiber lasers based on doped media with Er$^{3+}$, Ho$^{3+}$ and Dy$^{3+}$ ions have witnessed great progress to produce pulsed light in the range of 3-5 $\mu$m \cite{Jackson2012NP, Seddon2010OE}. However, due to the limited gain bandwidth, the mode-locked lasers usually deliver MIR light within a narrow spectral range \cite{Li2023OLT}. Notably, vibronic solid-state lasers in chalcogenide hosts offer a broad emission bandwidth, but are still confronting difficulties to fully cover the 2-5 $\mu$m MIR region with a single gain crystal \cite{Jackson2012NP}. Therefore, it is appealing to develop techniques for generating ultrashort MIR pulses over a broadband coverage.

In this context, the nonlinear frequency conversion approach offers a flexible and effective solution to generate broadly tunable coherent light in spectral regions inaccessible to existing lasers \cite{Dunn1999Science, Ren2021AS}. Compared to optical parametric generation (OPG) systems \cite{Nandy2020OL, Huang2021HPL, Murray2016OL, Ycas2018NP, Seidel2018SA}, synchronously pumped optical parametric oscillators (SPOPOs) typically offer a lower pump threshold, a higher conversion efficiency, and better spectral and spatial properties of the generated beams, thanks to the cavity power enhancement and spatial mode confinement \cite{Qin2015PJ, Chaitanya2014LPR}. These desirable features render the OPOs useful in many fields, such as high-precision molecular spectroscopy \cite{Long2023NP}, wide-field photothermal imaging \cite{Tamamitsu2024NP}, and non-invasive biomedical histopathology \cite{Junaid2019Optica}. Fueled by the great success in the aforementioned applications, there is on-going effort with an aim to promote OPO sources into practical scenarios beyond the laboratory environment.

To this end, there has been tremendous endeavor to leverage the fiber technology into conventional free-space OPOs for accessing desirable advantages of compactness, robustness, high beam quality, and flexible integration with external fiber systems \cite{Sudmeyer2001OL, Sudmeyer2004OL}. Compared to bulky laser systems, the integration of fiber components provides additional features like flexible configuration of cavity geometric length and free of tedious optical alignment \cite{Lamb2013OE, Yang2019OE, Donnell2020Optica, Alamgir2022OL}. In this direction, pioneering works have demonstrated so-called fiber-feedback OPOs (FOPOs), where a piece of optical fiber is used inside the cavity to replace the free-space beam path for the synchronous pumping operation \cite{Sudmeyer2001JPDAP}. Consequently, it is feasible to obtain powerful and tunable sub-picosecond pulses in a compact and stable fashion \cite{Steinle2016OL}. Such a hybrid free-space/fiber architecture is particularly attractive in constructing high-energy OPOs at a low repetition rate \cite{Kienle2010OL, Wu2022OL}. Indeed, most of optical paths in the resonator consist of the fiber, thus substantially reducing the complexity of operation and maintenance. Moreover, FOPOs have exhibited a strong immunity to intra-cavity loss and cavity-length drift, hence allowing a stable and robust operation without the need of active stabilization \cite{Donnell2020Optica, Alamgir2022OL, Sudmeyer2001JPDAP}. 

In previous works, a large pump intensity was commonly required to reach  parametric gains high enough to approach OPO threshold. The requirement of high-power pump becomes more critical at the presence of high-threshold FOPOs, particularly related to the cases of high repetition rates \cite{Ingold2015OL} or high intra-cavity losses \cite{Sudmeyer2001OL}. Additionally, the wavelength tuning of the FOPOs is typically achieved by changing the phase-matching conditions of the nonlinear crystal, for instance, relying on varying operating temperature or poling period \cite{Sudmeyer2004OL, Wu2023OE}. As a result, the spectral tuning speed is severely limited by the thermal equilibrium time or the mechanical translation speed. To fully reveal the potential of the FOPOs, it is highly demanded to develop novel techniques to reduce the pump threshold, extend the spectral coverage, and/or improve the tuning speed.

In this work, we have investigated two novel variants of FOPO systems. The first one is based on the insertion of a section of single-mode gain fiber into the resonator cavity, which can provide an additional gain for the circulating field in addition to the parametric gain. The synergic dual-gain operation results in a pumping threshold as small as 200 mW, which is much lower than that for previous demonstrations based on bulk nonlinear crystals. The relaxation of pump-power requirement makes it easier to access MIR pulsed light source in practice. In comparison to the single-pass configuration based on OPG, the implemented FOPO exhibits a much better pulse amplitude instability about 2.2\% even at a lower pumping power. The stable and robust operation is manifested by the large tolerance of cavity-length variation, which leads to a tight passive synchronization for the repetition rates of the involved pulse trains. Notably, the FOPO is constructed with polarization-maintaining fibers and elements, which favors long-term stability.Furthermore, we devise another FOPO system to achieve a widely tunable range based on a chirped-poling nonlinear crystal, which covers spectral ranges of 1350-1768 nm and 2450-4450 nm for the signal and idler beams, respectively. Thanks to the broad spectrum for the parametric gain, the wavelength tuning can be realized by simply adjusting the cavity length according to the group-velocity dispersion, thus eliminating the need of changing the phase-matching parameters. We believe that the presented FOPO sources would provide compact and flexible MIR sources for subsequent applications in material, chemical and biomedical fields.
 
\section{Experimental setup}
\begin{figure}[t!]
\centering
\includegraphics[width=0.95\columnwidth]{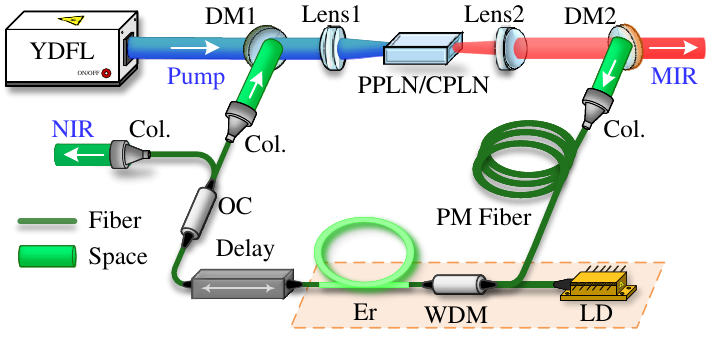}
\caption{Experimental setup for the fiber-feedback OPO at mid-infrared. A Yb-doped fiber laser (YDFL) at 1032 nm is used to pump the periodically-poled lithium niobate (PPLN) crystal for generating the signal and idler lights. The near-infrared signal beam is coupled into a section of polarization-maintaining (PM) fiber. After passing an Er-doped gain fiber, the signal is coupled back into the nonlinear crystal to form a ring cavity. Alternatively, the crystal can be replaced by a chirped-poling lithium niobate (CPLN) to access a broadband spectrum for the parametric gain. In this case, the gain fiber should be removed to fully exploit the wide tuning range. DM: dichroic mirror; Col.: collimator; WDM: wavelength division multiplexer; LD: laser diode.}
\label{fig1}
\end{figure}

Figure \ref{fig1} presents the schematic diagram of the FOPO, which consists of free-space and fiber sections in the optical cavity. The pump source originates from an ytterbium-doped fiber laser (YDFL, Langyan-tech, YbFemto ProH), which delivers ultrafast pulses at 1032 nm. The pulse duration is about 400 fs, and the repetition rate is 20.113 MHz. The pump laser outputs a maximum average power up to 1.5 W. A second-order nonlinear crystal based on periodically-poled lithium niobate (PPLN) is used to perform MIR parametric generation. The angular frequencies for the pump at $\omega_p$, the generated signal at $\omega_s$ and idler at $\omega_i$ satisfy the energy conservation law as $\omega_p =  \omega_s + \omega_i$. The PPLN crystal has multiple poling channels with a length of 20 mm. To approach the quasi-phase-matching condition, the crystal is placed in an oven with a temperature stabilization precision of 0.1 $^\circ$C. The generated MIR idler beam passes through a dichroic mirror, while the near-infrared (NIR) signal beam is reflected and coupled into a section of polarization-maintaining fiber (Nufern, PM1550-HP). Then the signal is sent into a 1-m Er-doped polarization-maintaining fiber (nLight, Er80-4/125-HD-PM) optically pumped by a laser diode (LD) at 976 nm. Finally, the signal is collimated back to the free space, and focuses into the nonlinear crystal to close the optical-path loop. Note that the lenses are achromatic to improve the conversion efficiency and coupling efficiency over a broad spectral band.

To implement the synchronous pumping, the cavity length of the FOPO can be tuned by the delay line (ZiguanTech, ZG-ODL-X-S9-1) inserted in the optical cavity. The delay line has a traveling range of 1.2 ns and a position precision of 50 fs. According to the group velocity dispersion of the fiber, the central wavelength of the circulating signal would be automatically adapted to match the pulse period of the pump excitation \cite{Becheker2018LPL}. Consequently, the signal pulse at each round would be temporally overlapped with the pump pulse. It is the involved parametric gain that sustains the circulation of the resonating signal pulse in the cavity. The signal is tapped for monitoring via the 1\% port of a fiber coupler.

\section{Results and discussion}
\begin{figure}[b!]
\centering
\includegraphics[width=0.9\columnwidth]{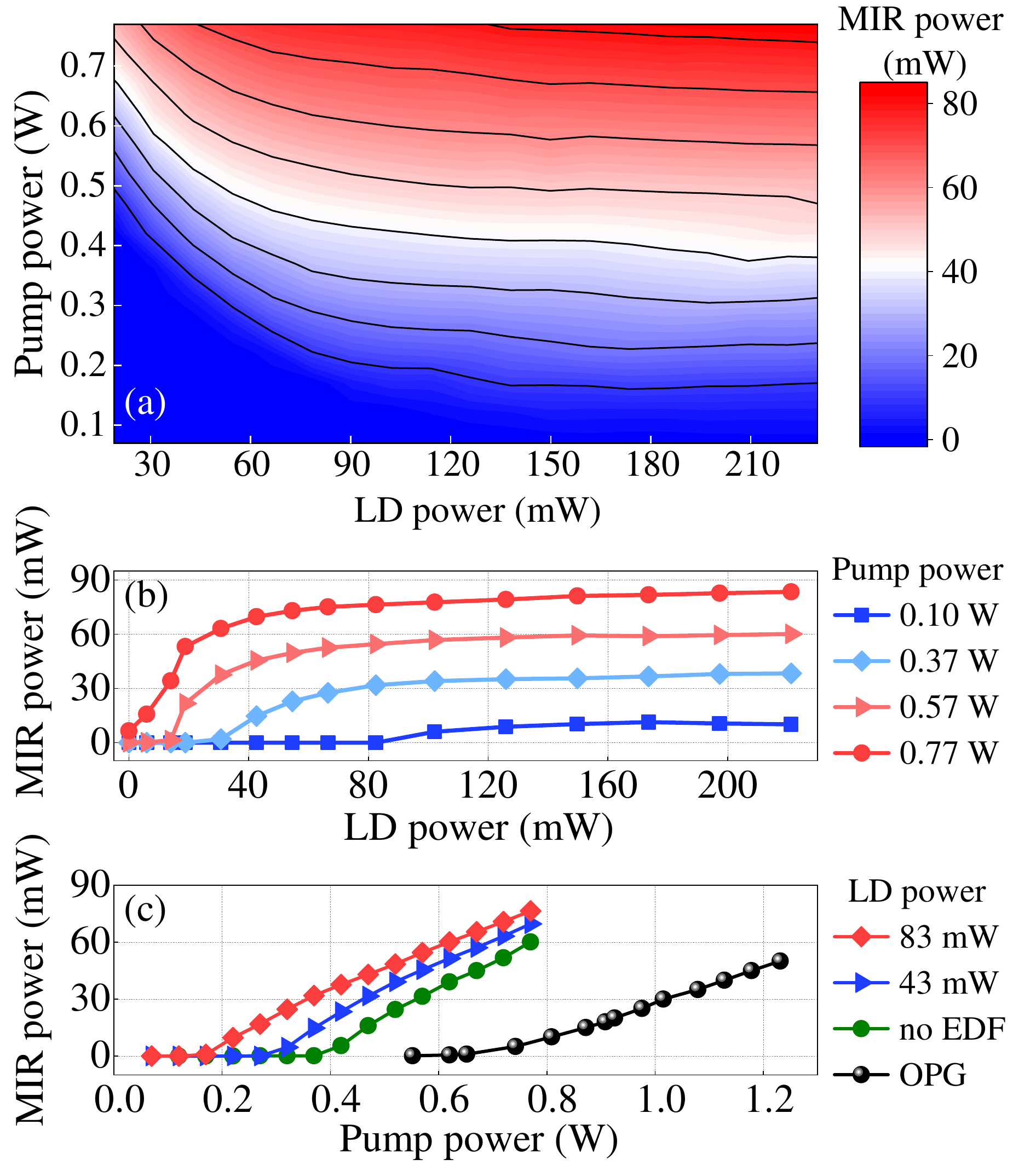}
\caption{(a) MIR power dependence on the LD and pump powers. (b) MIR power as a function of the LD power for the gain fiber in the presence of various pump powers for the nonlinear conversion. (c) MIR power versus the pump power for various LD powers. The behavior for single-pass configuration is given for comparison as denoted by OPG.}
\label{fig2}
\end{figure}

Now, we embark on a thorough examination of the performance for the FOPO. In contrast to previous instantiations \cite{Sudmeyer2001OL, Sudmeyer2004OL, Donnell2020Optica, Wu2022OL}, the presented scheme offers a doped-fiber gain in addition to the parametric gain, which could facilitate the reduction of the pumping threshold. Figure \ref{fig2}(a) presents the generated MIR power depending on the LD and pump powers. As shown in Fig. \ref{fig2}(b), the generated MIR power is investigated as increasing the LD power for the gain fiber at the presence of various pump power. Specifically, the MIR power tends to saturate at 83.5 mW at the pump power of 0.77 W, which results in an energy conversion efficiency about 32\% from the pump to the down-converted fields. For a lower pump power, the effect of fiber gain becomes more pronounced, where a higher LD power is required to obtain the MIR generation. At a pump power as low as 0.1 W, a saturated MIR power about 11.4 mW can be achieved at a LD power over 83 mW. It is thus seen that the fiber gain is essential in the case of a low parametric gain in the OPO. We note that the FOPO could deliver a higher power by resorting to more intense pumping, albeit that the effect of the fiber gain would become less prominent in such a high parametric gain regime.

\begin{figure}[b!]
\centering
\includegraphics[width=0.95 \columnwidth]{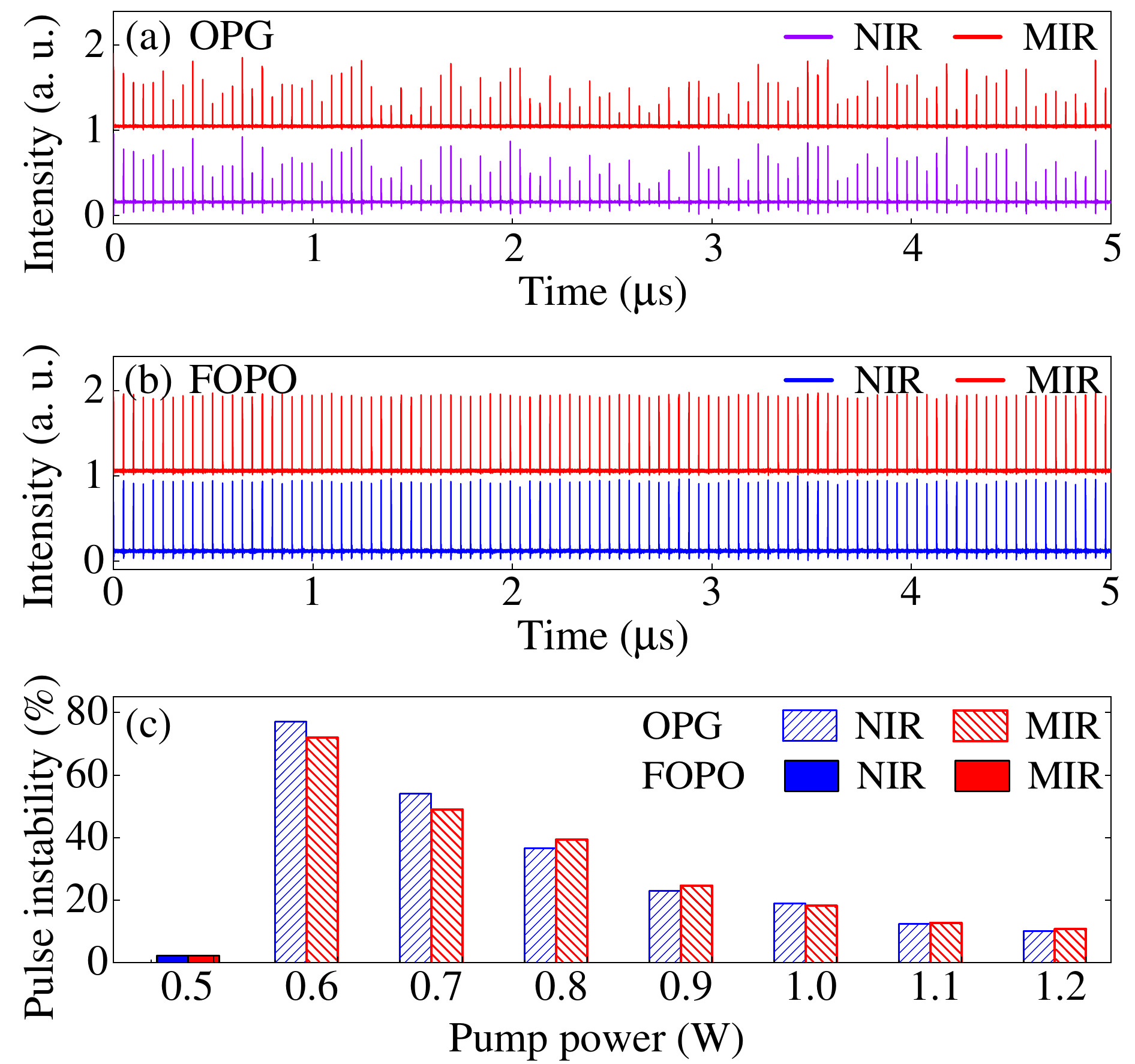}
\caption{Pulse instability at the FOPO and OPG configurations. Recorded pulse traces for the OPG (a) and FOPO (b) at pump powers of 0.8 and 0.5 W, respectively. (c) Measured pulse-to-pulse fluctuation for the near-infrared (NIR) signal and mid-infrared (MIR) idler pulses.}
\label{fig3}
\end{figure}

Similarly, Fig. \ref{fig2}(c) presents the MIR power dependence on the pump power at a fixed LD power. Typically, there is a threshold for the MIR generation, where the pump power should surpass a certain value to overcome the intra-cavity losses. In the case of LD powers of 83 and 43 mW, the pump thresholds are estimated to be about 0.2 and 0.3 W, respectively. In Fig. \ref{fig2}(c), the pump threshold is found to be about 0.4 W at the absence of the gain fiber. Hence, the involved supplementary fiber gain renders it possible to reduce the OPO threshold. Indeed, the additional fiber gain is provided by additional power from the LD. But the LD pump is more cost-effective, and only requires a continuous-wave operation, which differs from the ultrashort pulse requirement for the parametric pumping in SPOPOs. In this sense, the LD power is not taken into the account for evaluating the pump threshold for the FOPO. As a direct comparison, the pump threshold is identified to be about 0.6 W for the single-pass configuration in the regime of OPG. The demonstrated threshold here is much lower than reported values at the watt level for previous FOPOs with only the parametric gain \cite{Sudmeyer2001OL, Sudmeyer2004OL, Kienle2010OL, Wu2022OL}. Such a dual-gain configuration is favorable for MIR generation at the presence of low-power pump sources. It is worth noting that recent advances on monolithic OPOs have demonstrated a substantial reduction of the pump threshold based on photonic waveguides \cite{Langrock2007OL, Jornod2023Optica} or microring resonators \cite{Lu2021Optica, Bruch2019Optica, Hwang2023Optica}. However, it is still challenging for the chip-scale OPOs to extend the operation wavelength into the MIR spectral region due to the technical difficulties on spatial mode matching and dispersion matching for the involved disparate and tunable wavelengths.

\begin{figure}[b!]
\centering
\includegraphics[width=1 \columnwidth]{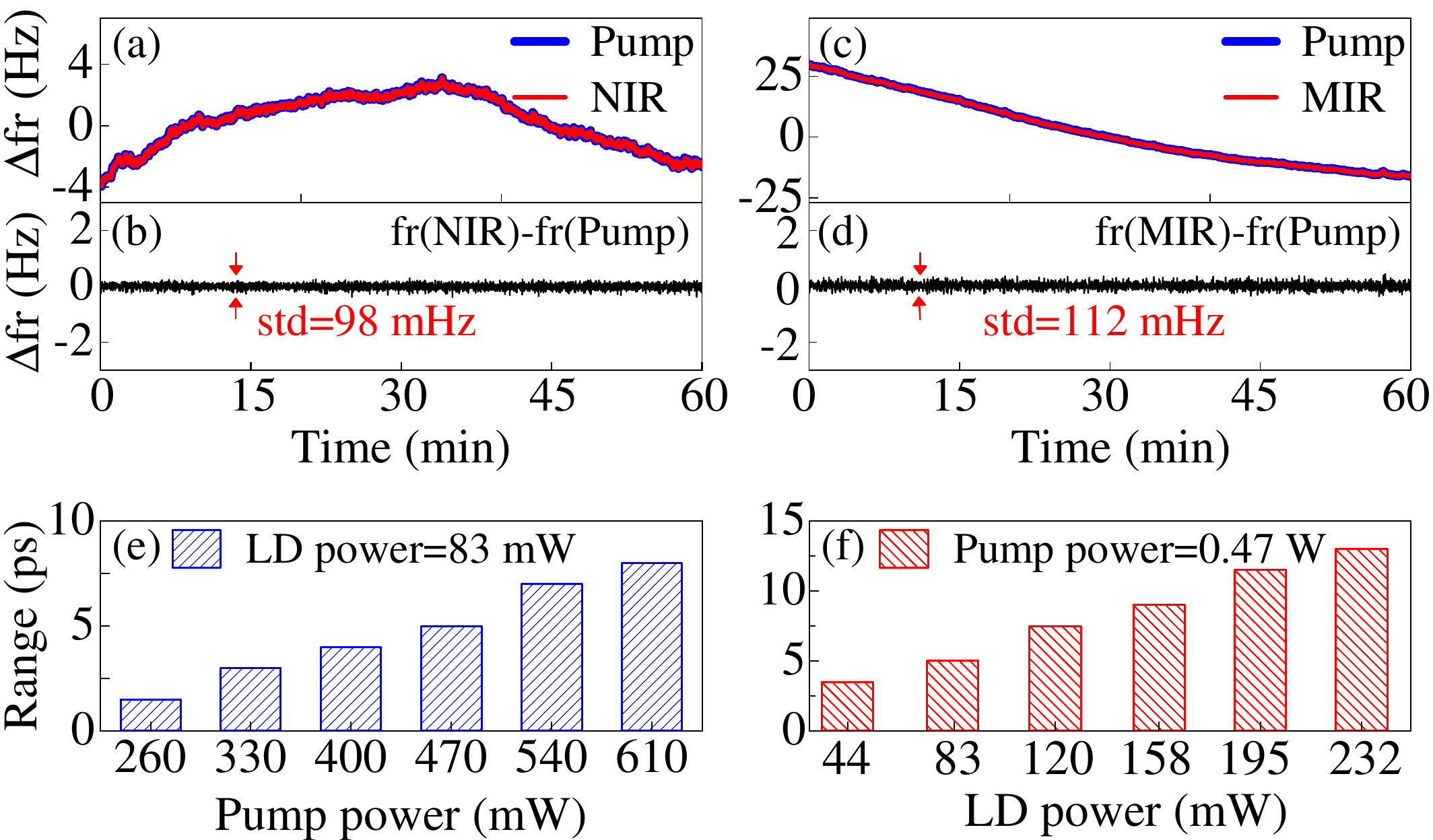}
\caption{Passive synchronization between the pump, signal and idler pulses. (a) Measured repetition rates for the pump and signal pulses during a period of 1 hour. The repetition rate is offset by 20.113 MHz for clarity. (b) Relative repetition rate for the synchronized pulse trains, indicating a standard deviation of 98 mHz. (c) Evolution of repetition rates for the pump and idler pulses, and for the corresponding differential values (d). (e, f) Tolerance range for obtaining synchronized pulses versus the pump power (e) and LD power (f).} 
\label{fig4}
\end{figure}

\begin{figure*}[t!]
\centering
\includegraphics[width=0.8\textwidth]{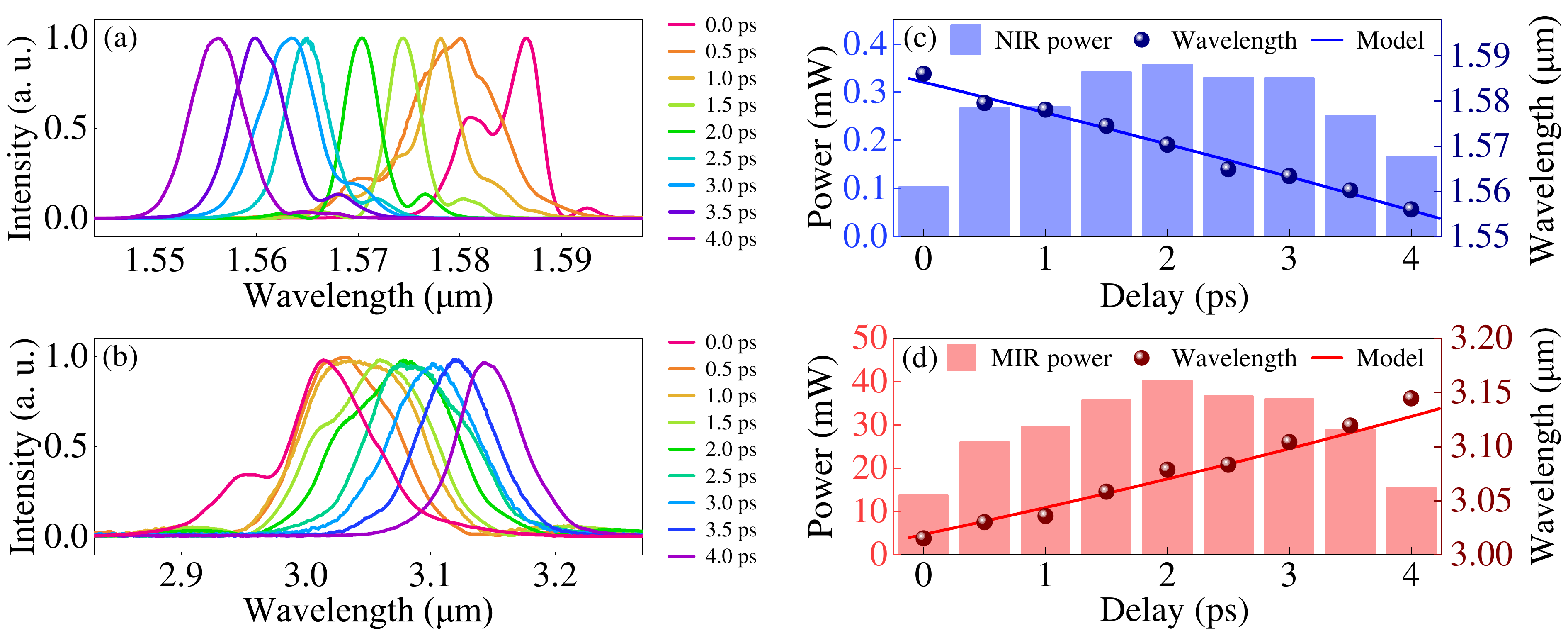}
\caption{Generated spectra and output power as a function of the cavity-length variation. (a, b) Spectral evolution for the signal (a) and idler (b) beams as tuning the delay line in the OPO cavity. (c, d) Central wavelengths and output power as a function of the delay for the signal (c) and idler (d) beams. The solid lines are calculated from the theoretical model.}
\label{fig5}
\end{figure*}

Next, we turn to characterize the pulse stability for the MIR source. Besides a lower pumping threshold, the OPO usually features a more stable pulse generation with a comparison to the single-pass scheme, particularly at a modest pumping intensity. Figures \ref{fig3}(a) and (b) present the recorded pulse trains for the OPG and FOPO at pump power of 0.8 and 0.5 W, respectively. The signal and idler traces for the OPG show pulse-to-pulse amplitude fluctuations of 37$\%$ and 39$\%$, while the pulse fluctuations for the FOPO traces are 2.2$\%$ and 2.3$\%$. As shown in Fig. \ref{fig3}(c), the pulse instability in OPG scheme can be generally improved by increasing the pump power. At a 1.2-W pump, the signal and idler pulse instabilities are measured to be 10$\%$ and 11$\%$, which are still worse than the achieved values for the FOPO at a lower pump power. Note that a stable MIR train can also be obtained via the difference-frequency generation (DFG) \cite{Huang2021HPL, Murray2016OL}. In this scenario, a synchronous seed laser is typically required to induce the parametric conversion process, which inevitably increases the system complexity. Instead, the down-converted signal here is recycled to serve as the seeding, thus providing an effective and compact way to generate stable MIR pulses.

In the following, we aim to investigate the temporal synchronization among the pump, signal, and idler pulses. Figure \ref{fig4}(a) gives the repetition rates for the pump and signal pulses, which are measured by a frequency counter (Tektronix, FCA3100). The sampling time for the counter is set to be 1 s, and the acquisition time is one hour. Since the cavity length of the pump laser is not actively stabilized, the repetition rate thus fluctuates over time due to the ambient perturbation or thermal variation. However, the relative repetition rate $\Delta \text{fr}$ between the pump and signal pulses is kept constant as shown in Fig. \ref{fig4}(b). The standard deviation of $\Delta \text{fr}$  is calculated to be 98 mHz, which corresponds to a relative frequency fluctuation of 4.9$\times$10$^{-9}$. Identically, the repetition rates for the pump and idler are recorded as given in Fig. \ref{fig4}(c), while the differences are presented in Fig. \ref{fig4}(d). We note that the measured repetition-rate uncertainties are close to the system instability for the used frequency counter, which indicates a tight temporal synchronization among the three involved beams in the FOPO. Furthermore, the tolerance range for the passive synchronization is found to be linearly dependent on the pump and LD powers, as illustrated in Figs. \ref{fig4}(e) and (f), respectively. Such a passive synchronization becomes attractive by eliminating the active locking of the repetition rates for the involved pulse trains. The generated multi-color synchronous pulses could be useful in various applications, such as complementary vibrational spectroscopy \cite{Hashimoto2019NC} and upconversion infrared imaging \cite{Junaid2019Optica}.  

\begin{figure*}[t!]
\centering
\includegraphics[width=0.8 \textwidth]{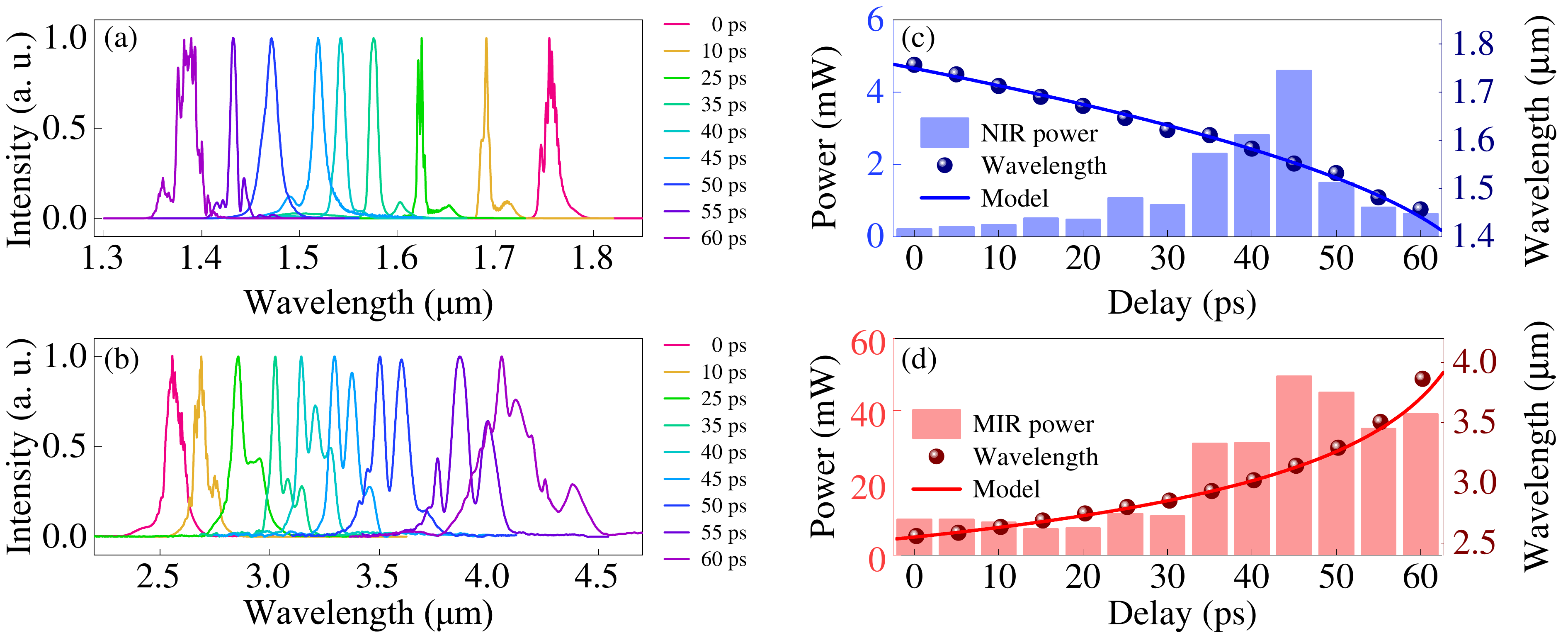}
\caption{Recorded optical spectra for the signal (a) and idler (b) light as the tuning delay line in the case of a CPLN nonlinear crystal used in the OPO cavity. (c, d) Central wavelengths and output power as a function of the delay for the signal (c) and idler (d) beams. The solid lines are calculated from the theoretical model.}
\label{fig6}
\end{figure*}
 
In the experiment, we further investigate the underlying mechanism for the passive synchronization with a large tolerance range. The  poling period for the PPLN crystal is 30.5 $\mu$m, and the operation temperature is set to be 27 $^\circ$C. The pump power and LD power are fixed at 0.47 W and 83 mW, respectively. As shown in Fig. \ref{fig5}(a), the center wavelength of the signal spectrum decreases as the cavity length increases. In a longer cavity, the generated signal should travel faster to catch up the pump pulse for obtaining sufficient parametric gain. As a result, the signal spectrum would exhibit a blue shift due to the net anomalous dispersion for the intra-cavity fibers. Specifically, the group velocity dispersions for the single-mode fiber and Er-doped gain fiber are specified to be  about 19 and -23 ps/nm/km at 1570 nm, respectively. The lengths of the two type of fibers are 8.3 m and 1 m, which result in a total dispersion of 0.13 ps/nm. As shown in Fig. \ref{fig5}(c), the induced spectral shift agrees well with the theoretical expectation. Notably, the required time delay to tune the central wavelength by an amount of a typical bandwidth of 6 nm is about 0.8 ps, which corresponds to a cavity-length variation of 240 $\mu$m in the free space. Such a change is much larger than length fluctuation for a free-running OPO cavity. As a result, a stable signal spectrum can be obtained when the delay line is fixed at a specific position. In the signal-resonating OPO, the amplified spontaneous emission (ASE) from the LD pumping in the gain fiber imposes a negligible effect on the measured spectrum. Indeed, only the signal wave with a specific spectro-temporal property can oscillate in the cavity. More specifically, its spectrum should match the bandwidth of phase-matching window of the nonlinear crystal, and the temporal profile should be overlapped with the ultrashort pulse.

In comparison to the  free-space OPOs, the involved fibers in the FOPO introduce a large intra-cavity dispersion, thus allowing a wide tolerance range for the cavity-length variation \cite{Becheker2018LPL}. Once the signal wavelength is determined, the MIR idler wavelength can be inferred from the energy conservation law. Figure \ref{fig5}(b) gives the recorded MIR spectra for various time delays. The central wavelength for the idler beam increases as elongating the cavity length as shown in Fig. \ref{fig5}(d). The output average powers for the signal and idler are given in Figs. \ref{fig5}(c) and (d), respectively. Note that the spectral tuning range of 1553-1586 nm for the generated signal is mainly determined by the gain window for the Er-doped gain fiber. The pulse duration of the signal is measured to be about 380 fs by an optical autocorrelator (APE, Pulsecheck SM2000). The pulse duration of the idler beam is inferred to be about 390 fs from the cross-correlation trace based on the sum-frequency generation between two down-converted beams.

Finally, we propose and implement a widely tunable MIR FOPO system by using a chirped-poling lithium niobate (CPLN) crystal. The CPLN has a ramping period from 25 to 32 $\mu$m along a length of 20 mm, which allows an idler phase-matching window from 2 to 5 $\mu$m. In this configuration, the Er-doped gain fiber is removed to accommodate the broadband spectral range for the NIR signal. The pump power increases to 0.8 W due to the absence of fiber gain and the lower parametric gain for the CPLN crystal. Notably, the CPLN crystal can operate at room temperature, which excludes the oven as used for the PPLN crystal. As depicted in Fig. \ref{fig6}(a), the tolerance for the time delay is extended to over 60 ps, which results in a wide tuning range for the signal beam from 1350 to 1768 nm. Correspondingly, the spectrum for the MIR idler is given in Fig. \ref{fig6}(b), which indicates a spectral coverage from 2450 to 4450 nm. The spectral profiles and bandwidths vary at different time delays, which are governed by the phase-matching condition at each adapted poling period. The central wavelengths for the generated signal and idler are in good agreement with the theoretical predictions as shown in Fig. \ref{fig6}(c) and (d), respectively. It is noteworthy that the broad tuning range is achieved by only tuning the time delay, which contrasts to previous demonstrations that rely on a synergic operation on adapting the phase-matching condition for the nonlinear crystal \cite{Sudmeyer2004OL, Wu2023OE}. Indeed, for a conventional PPLN crystal, the parametric gain spectrum is typically altered by changing the operating temperature or poling period, which would inevitably limit the spectral tuning speed. Here, the intrinsically broadband profile of the parametric gain makes it possible to tune the output wavelength in an agile fashion. For example, high-speed optical delay lines are commercially available, which allow a 20-Hz delay modulation over a traveling range of 60 ps (MesaPhotonics, ODL). Such a travel delay stage offers the possibility to facilitate a high-speed spectral sweeping rate up to 31,000 nm/s, which would be highly demanded for real-time MIR hyperspectral imaging \cite{Junaid2019Optica}.

\section{Conclusion}
In summary, we have devised and implemented two novel variants of the FOPO system, which aim to reduce the pump threshold and extend the spectral coverage. Particularly, the introduction of additional fiber gain in the cavity enables the MIR generation at a low pump power. In the experiment, the resonating signal spectrum is restricted by the C-band gain profile of erbium-doped fiber. The operation wavelength is possible to cover the O-band region by using a bismuth-doped gain fiber, which allows to access more flexible MIR wavelengths for the idler beam. Furthermore, we have investigated a widely tunable FOPO based on a CPLN crystal, which leads to the preparation of a MIR ultrafast source from 2450 to 4450 nm. Thanks to the intrinsically broad spectral profile for the parametric gain, the wavelength sweeping can be performed by simply tuning the cavity length, without the need for adapting the phase-matching condition as required in previous demonstrations based on periodically poled nonlinear crystals \cite{Sudmeyer2004OL, Wu2023OE}. 

Currently, the pulse durations for the generated signal and idler beams are at the sub-picosecond level, which can be further optimized to the femtosecond regime with the help of intra-cavity dispersion management \cite{Donnell2020Optica}. In combination with advanced photonic technology, high-integration FOPOs could be envisioned by using emerging nonlinear platforms based on waveguides \cite{Langrock2007OL}, thin films \cite{Lu2021Optica, Hwang2023Optica}, or microring resonators \cite{Bruch2019Optica}. Therefore, we believe that the presented paradigms and achieved performances of the FOPOs would stimulate a variety of subsequent MIR applications, such as infrared sensing, material inspection, and molecular spectroscopy.

\vspace{6pt}
\noindent  {\fontfamily{phv}\selectfont 
\normalsize \textbf{Funding.} 
}
\noindent This work was supported by National Natural Science Foundation of China (62175064, 62235019, 62035005); Shanghai Pilot Program for Basic Research (TQ20220104); Natural Science Foundation of Chongqing   (CSTB2023NSCQ-JQX0011, CSTB2022NSCQ-MSX0451, CSTB2022TIAD-DEX0036); Shanghai Municipal Science and Technology Major Project (2019SHZDZX01); Fundamental Research Funds for the Central Universities.

\vspace{6pt}
\noindent  {\fontfamily{phv}\selectfont 
\normalsize \textbf{Disclosures.} 
}
\noindent The authors declare no conflict of interest.

\vspace{6pt}
\noindent  {\fontfamily{phv}\selectfont 
\normalsize \textbf{Data availability.} 
}
\noindent The data that support the findings of this study are available from the corresponding author upon reasonable request.

\vspace{6pt}
\noindent  {\fontfamily{phv}\selectfont 
\normalsize \textbf{Keywords.} 
}
\noindent  Optical parametric oscillator, Mid-infrared generation, Fiber laser, Nonlinear optics


\begin{thebibliography}{100}

\bibitem{Vodopyanov2020Book} K. L. Vodopyanov, Laser-Based Mid-Infrared Sources and Applications, John Wiley \& Sons (2020).

\bibitem{Schliesser2012NP} A. Schliesser, N. Picqu\'{e}, and T. W. H\"{a}nsch, ``Mid-infrared frequency combs," Nat. Photon. \textbf{6}, 440 (2012).

\bibitem{Hashimoto2019NC}  K. Hashimoto, V. R. Badarla, A. Kawai, and T. Ideguchi, ``Complementary vibrational spectroscopy," Nat. Commun. \textbf{10}, 4411 (2019).

\bibitem{Yao2012NP} Y. Yu, A. J. Hoffman, C. F. Gmachl,  ``Mid-infrared quantum cascade lasers," Nat. Photon. \textbf{6}, 432 (2012).

\bibitem{Jackson2012NP} S. D. Jackson, ``Towards high-power mid-infrared emission from a fibre laser," Nat. Photon. \textbf{6}, 423 (2012).

\bibitem{Seddon2010OE} A. B. Seddon, Z. Tang, D. Furniss, S. Sujecki, and T. M. Benson, ``Progress in rare-earth-doped mid-infrared fiber lasers," Opt. Express \textbf{18}, 26704 (2010).

\bibitem{Li2023OLT} X. Li, X. Huang, X. Hu, X. Guo, and Y. Han, ``Recent progress on mid-infrared pulsed fiber lasers and the applications," Opt. Laser Technol. \textbf{158}, 108898 (2023). 

\bibitem{Dunn1999Science} M. H. Dunn, M. Ebrahimzadeh, ``Parametric Generation of Tunable Light from Continuous-Wave to Femtosecond Pulses," Science \textbf{286}, 1513 (1999).

\bibitem{Ren2021AS} T. Ren, C. Wu, Y. Yu, T. Dai, F. Chen, Q. Pan. ``Development Progress of 3-5 $\mu$m Mid-Infrared Lasers: OPO, Solid-State and Fiber Laser," Appl. Sci. \textbf{11}, 11451 (2021).

\bibitem{Nandy2020OL} B. Nandy, S. C. Kumar, and M. Ebrahim-Zadeh, ``Fiber laser pumped high repetition rate picosecond optical parametric generation and amplification in MgO:PPLN," Opt. Lett. \textbf{45}, 6126 (2020).

\bibitem{Huang2021HPL} K. Huang, Y. Wang, J. Fang, H. Chen, M. Xu, Q. Hao, M. Yan, and H. Zeng, ``Highly efficient difference-frequency generation for mid-infrared pulses by passively synchronous seeding," High Power Laser Sci. Eng. \textbf{9}, e4 (2021).

\bibitem{Murray2016OL} R. T. Murray, T. H. Runcorn, E. J. R. Kelleher, and J. R. Taylor, ``Highly efficient mid-infrared difference-frequency generation using synchronously pulsed fiber lasers," Opt. Lett. \textbf{41}, 2446 (2016).

\bibitem{Ycas2018NP} G. Ycas, F. R. Giorgetta, E. Baumann, I. Coddington, D. Herman, S. A. Diddams, N. R. Newbury, ``High-coherence mid-infrared dual-comb spectroscopy spanning 2.6 to 5.2 $\mu$m," Nat. Photon. \textbf{12}, 202 (2018). 

\bibitem{Seidel2018SA} M. Seidel, X. Xiao, S. A. Hussain, G. Arisholm, A. Hartung, K. T. Zawilski, P. G. Schunemann, F. Habel, M. Trubetskov, V. Pervak, O. Pronin, and F. Krausz, ``Multi-watt, multi-octave, mid-infrared femtosecond source," Sci. Adv. \textbf{4}, eaaq1526 (2018).

\bibitem{Qin2015PJ} Z. Qin, G. Xie, W. Ge, P. Yuan and L. Qian, ``Over 20-W Mid-Infrared Picosecond Optical Parametric Oscillator," IEEE Photon. J. \textbf{7}, 1 (2015).

\bibitem{Chaitanya2014LPR} S. C. Kumar, A. Esteban-Martin, T. Ideguchi, M. Yan, S. Holzner, T.W. H\"{a}nsch, N. Picqu\'{e}, M. Ebrahim-Zadeh,``Few-cycle, broadband, mid-infrared optical parametric oscillator pumped by a 20-fs Ti:sapphire laser," Laser Photon. Rev. \textbf{5}, 86 (2014).

\bibitem{Long2023NP} D. A. Long, M. J. Cich, C. Mathurin, A. T. Heiniger, G. C. Mathews, A. Frymire, and G. B. Rieker, ``Nanosecond time-resolved dual-comb absorption spectroscopy," Nat. Photon. \textbf{18}, 127 (2023).

\bibitem{Tamamitsu2024NP} M. Tamamitsu, K. Toda, M. Fukushima, V. R. Badarla, H. Shimada, S. Ota, K. Konishi, and T. Ideguchi, ``Mid-infrared wide-field nanoscopy," Nat. Photon. arXiv:2306.08245 (2024).

\bibitem{Junaid2019Optica} S. Junaid, S. C. Kumar, M. Mathez, M. Hermes, N. Stone, N. Shepherd, M. Ebrahim-Zadeh, P. Tidemand-Lichtenberg, and C. Pedersen, ``Video-rate, mid-infrared hyperspectral upconversion imaging," Optica \textbf{6}, 702 (2019).

\bibitem{Sudmeyer2001OL} T. S\"{u}dmeyer, J. A. der Au, R. Paschotta, U. Keller, P. G. R. Smith, G. W. Ross, and D. C. Hanna, ``Femtosecond fiber-feedback optical parametric oscillator," Opt. Lett. \textbf{26}, 304 (2001).

\bibitem{Sudmeyer2004OL} T. S\"{u}dmeyer, E. Innerhofer, F. Brunner, R. Paschotta, T. Usami, H. Ito, S. Kurimura, K. Kitamura, D. C. Hanna, and U. Keller, ``High-power femtosecond fiber-feedback optical parametric oscillator based on periodically poled stoichiometric LiTaO$_{3}$," Opt. Lett. \textbf{29}, 1111 (2004).

\bibitem{Lamb2013OE} E. S. Lamb, S. Lefrancois, M. Ji, W. J. Wadsworth, X. S. Xie, and F. W. Wise, ``Fiber optical parametric oscillator for coherent anti-Stokes Raman scattering microscopy," Opt. Lett. \textbf{38}, 4154 (2013).

\bibitem{Yang2019OE} K. Yang, S. Zheng, P. Ye, Q. Hao, K. Huang, and H. Zeng, ``Fiber-based optical parametric oscillator with flexible repetition rates by rational harmonic pumping," Opt. Express \textbf{27}, 4897 (2019).

\bibitem{Donnell2020Optica} C. F. O' Donnell, S. C. Kumar, T. Paoletta, and M. Ebrahim-Zadeh, ``Widely tunable femtosecond soliton generation in a fiber-feedback optical parametric oscillator," Optica \textbf{7}, 426 (2020)

\bibitem{Alamgir2022OL} I. Alamgir, M. Rezaei, and M. Rochette, ``Fiber optical parametric oscillator made of soft glass," Opt. Lett. \textbf{47}, 3451 (2022).

\bibitem{Sudmeyer2001JPDAP} T. S\"{u}dmeyer, J. A. der Au, R. Paschotta, U. Keller, P. G. R. Smith, G. W. Ross, and D. C. Hanna, ``Novel ultrafast parametric systems: high repetition rate single-pass OPG and fibre-feedback OPO," J. Phys. D. Appl. Phys. \textbf{34}, 2433 (2001).

\bibitem{Steinle2016OL} T. Steinle, F. Morz, A. Steinmann, and H. Giessen, ``Ultra-stable high average power femtosecond laser system tunable from 1.33 to 20 $\mu$m," Opt. Lett. \textbf{41}, 4863 (2016).

\bibitem{Kienle2010OL} F. Kienle, P. S. Teh, S. Alam, C. B. E. Gawith, D. C. Hanna, D. J. Richardson, and D. P. Shepherd, ``Compact, high-pulse-energy, picosecond optical parametric oscillator," Opt. Lett. \textbf{35}, 3580 (2010).

\bibitem{Wu2022OL} Y. Wu, S. Liang, Q. Fu, T. D. Bradley, F. Poletti, D. J. Richardson, and L. Xu, ``High-energy, mid-IR, picosecond fiber-feedback optical parametric oscillator," Opt. Lett. \textbf{47}, 3600 (2022).

\bibitem{Ingold2015OL} K. A. Ingold, A. Marandi, M. J. F. Digonnet, and R. L. Byer, ``Fiber-feedback optical parametric oscillator for half-harmonic generation of sub-100-fs frequency combs around 2 $\mu$m," Opt. Lett. \textbf{40}, 4368 (2015).

\bibitem{Wu2023OE} Y. Wu, Q. Fu, S. Liang, F. Poletti, D. J. Richardson, and L. Xu, ``15-$\mu$J picosecond hollow core fiber feedback optical parametric oscillator," Opt. Express \textbf{31}, 23419 (2023).

\bibitem{Becheker2018LPL} R. Becheker, M. Tang, P. Hanzard, A. Tyazhev, A. Mussot, A. Kudlinski, A. Kellou, J. Oudar, T. Godin, and A. Hideur, ``High-energy dissipative soliton-driven fiber optical parametric oscillator emitting at 1.7-$\mu$m," Phys. Lett. \textbf{15}, 115103 (2018).

\bibitem{Langrock2007OL} C. Langrock and M. M. Fejer, ``Fiber-feedback continuous-wave and synchronously-pumped singly-resonant ring optical parametric oscillators using reverse-proton-exchanged periodically-poled lithium niobate waveguides," Opt. Lett. \textbf{32}, 2263 (2007).

\bibitem{Jornod2023Optica} N. Jornod, M. Jankowski, L. M. Kr\"{u}ger, V. J. Wittwer, N. Modsching, C. Langrock, C. R. Phillips, U. Keller, T. S\"{u}dmeyer, and M. M. Fejer, ``Monolithically integrated femtosecond optical parametric oscillators," Optica \textbf{10}, 826 (2023).

\bibitem{Lu2021Optica}  J. Lu, A. Al Sayem, Z. Gong, J. B. Surya, C. L. Zou, and H. X. Tang, ``Ultralow-threshold thin-film lithium niobate optical parametric oscillator," Optica \textbf{8}, 539 (2021). 

\bibitem{Bruch2019Optica} A. W. Bruch, X. Liu, J. B. Surya, C. Zou, and H. X. Tang, ``On-chip ${\chi^{(2)}}$ microring optical parametric oscillator," Optica \textbf{6}, 1361 (2019).

\bibitem{Hwang2023Optica} A. Y. Hwang, H. S. Stokowski, T. Park, M. Jankowski, T. P. McKenna, C. Langrock, J. Mishra, V. Ansari, M. M. Fejer, and A. H. Safavi-Naeini, ``Mid-infrared spectroscopy with a broadly tunable thin-film lithium niobate optical parametric oscillator," Optica \textbf{10}, 1535 (2023).

  
\end{thebibliography}
\end{document}